\documentclass[11pt,a4paper]{article}
\usepackage{jcappub}
\usepackage{color}
\usepackage{amssymb,amsmath,amsfonts}
\usepackage{latexsym}
\usepackage[totalwidth=17cm,totalheight=23.8cm]{geometry}
\usepackage{hyperref}
\usepackage{graphicx}   
\usepackage{verbatim}   
\usepackage{color}      
\usepackage{hyperref}   
\usepackage{float}
\usepackage{amssymb}
\usepackage[sort&compress]{natbib}
\def\rmi{{\rm i}}

\title{Quantum cosmological consistency condition for inflation}

\author[a]{Gianluca Calcagni,}
\author[b]{Claus Kiefer,}
\author[c]{Christian F. Steinwachs}

\affiliation[a]{Instituto de Estructura de la Materia, CSIC,\\
calle Serrano 121, 28006 Madrid, Spain}
\affiliation[b]{Institut f\"ur Theoretische Physik, Universit\"at zu K\"oln,\\
Z\"ulpicher Strasse 77, 50937 K\"oln, Germany}
\affiliation[c]{Physikalisches Institut, Albert-Ludwigs-Universit\"at Freiburg,\\
Hermann-Herder-Str.~3, 79104 Freiburg, Germany}

\emailAdd{calcagni@iem.cfmac.csic.es}
\emailAdd{kiefer@thp.uni-koeln.de}
\emailAdd{christian.steinwachs@physik.uni-freiburg.de}

\abstract{

\noindent We investigate the quantum cosmological tunneling scenario for
inflationary models. Within a path-integral approach, we
derive the corresponding tunneling probability distribution. A sharp
peak in this distribution can be interpreted as the initial condition
for inflation and therefore as a quantum cosmological prediction for
its energy scale. This energy scale is also a genuine prediction of
any inflationary model by itself, as the primordial gravitons
generated during inflation leave their imprint in the $B$-polarization
of the cosmic microwave background. 
In this way, one can derive a consistency condition for inflationary models that
guarantees compatibility with a tunneling origin and can 
lead to a testable quantum cosmological prediction. The
general method is demonstrated explicitly for the model of natural
inflation. 
}

\arxivnumber{1405.6541}

\begin{document}

\maketitle
\flushbottom

\section{Introduction}
Cosmology without the mechanism of inflation (see e.g.\ \cite{Mukhanov:book}) 
seems inconceivable. Not only does inflation solve many conceptual
problems of the old hot big-bang theory, it is also in excellent
agreement with experimental data of ever increasing precision
\cite{Komatsu:2010fb, Hinshaw:2012aka, Ade:2013zuv,Ade:2013uln}. In
fact, it is hard to devise a mechanism different from inflation that
could solve all cosmological obstacles and, at the same time, does not
contradict any observation. 

Despite all its success, however, the nature and the origin of
inflation remain unexplained so far. Most inflationary models are
based on a scalar field $\varphi$ ---the inflaton. Its
identification with a scalar field in particle-physics models is still
debated. For instance, in the model of non-minimal Higgs inflation,
$\varphi$ has been identified with the observed Standard Model Higgs
boson
\cite{Bezrukov:2007ep,Barvinsky:2008ia,Bezrukov:2008ej,DeSimone:2008ei,Barvinsky:2009fy,Bezrukov:2009db,Bezrukov:2010jz,Barvinsky:2009ii,
  Allison:2013uaa}.\footnote{See also variants of Higgs inflation such
  as the `new Higgs inflation' model, where a minimally coupled
  scalar field with a non-canonical kinetic term coupled to the
  Einstein tensor was considered 
  \cite{Germani:2010gm,Germani:2010ux,Germani:2014hqa}.} However, even
in scenarios where $\varphi$ is embedded in a realistic theory, the
question of why inflation has started in the first place remains
mostly unresolved.

Inflation presupposes an already pre-existing classical background on
which tiny primordial quantum fluctuations can propagate
\cite{Starobinsky1979,Mukhanov:1981xt,Hawking:1982cz,Guth:1982ec,Starobinsky:1982ee,Bardeen:1983qw}. 
During inflation, these quantum fluctuations experience an effective
quantum-to-classical transition; see
\cite{PS96,Kiefer:2006je,Kiefer:2008ku} and the references therein.

At the most fundamental level, a classical background does not exist. The reason is that,
given the fundamental quantum character of all matter interactions, it
is expected that gravity (and therefore spacetime) has to be quantized as well. 
There are many different approaches to quantum gravity
\cite{Ori09,Kiefer:2012boa,Kiefer:2013jqa}.
In the traditional canonical approach, when restricting to cosmology,
we obtain the Wheeler--DeWitt equation that governs the quantum dynamics of the universe.
This is a differential equation which has to be accompanied by a proper boundary condition. 
In the absence of a fully developed theory of quantum gravity,
cosmological considerations offer at least a heuristic guideline for
natural choices of such boundary conditions. 
The hope is that this will ultimately lead to observable quantum
cosmological effects
\cite{Calcagni:2012vw,KK12a,KK12b,Bini:2013fea,Sasha14}.  There also exists a 
path-integral analogue to the canonical approach, which will be used here.

Two of the most influential proposals for boundary conditions
are the so called
no-boundary \cite{Hawking:1982,Hartle:1983ai}  and tunneling
\cite{Vilenkin:1982de,Vilenkin:1983xq,Linde:1983mx, Vilenkin:1984wp,
  Rubakov:1984ki, Zeldovich:1984vk} conditions; see, for example,
\cite{Kiefer:2012boa} for a review. 
The underlying picture behind the tunneling condition is that our
universe as a whole was created by a `quantum tunneling from
nothing'. (As we shall briefly discuss below, however, this picture can
at best be seen as a metaphor.) 
The tunneling condition seems to be preferred over the no-boundary
condition, in the sense that it can lead to a successful
post-nucleating phase of inflation; see  \cite{Barvinsky:2009jd} and
references therein.

It is, however, not self-evident that an inflationary model and the
tunneling process can always be combined into one consistent
scenario. Typically, the tunneling proposal is believed to give rise
to a sustainable inflationary phase because it predicts a conditional
probability of the field values peaked at large values of the potential. This accords with
chaotic inflation where, in its simplest incarnation, the potential is
a monomial of $\varphi$.\footnote{On the other hand, the no-boundary
  proposal can accommodate inflation after introducing a re-weighting
  of the probability \cite{HHH1,HHH2,HHH3,Hartle:2010dq}.} 

The purpose of this paper is to derive a compatibility condition testing whether the origin of inflationary models favored by the 2013
\textsc{Planck} release \cite{Ade:2013uln, Martin:2013gra,
  Tsujikawa:2013ila} can consistently be explained by a quantum
tunneling of the universe. The requirement of consistency then might
lead to restrictions for the parameters of the underlying inflationary
models and therefore to testable quantum cosmological predictions. For completeness, a separate confrontation with \textsc{Bicep2} results \cite{Ade:2014xna} is also carried out, regardless of the ongoing debate on their ultimate validity \cite{MoSe}.

The paper is structured as follows. In section \ref{II}, we introduce
the formalism of the Euclidean instanton and construct the quantum
cosmological tunneling distribution. 
In section \ref{III}, we consider the model of natural inflation and
derive a consistency condition for the tunneling scenario. 
We conclude in section \ref{IV} by summarizing our results and comment
on a similar analysis for different models of inflation as well as on
a more ambitious quantum analysis. 

\section{Quantum origin of the cosmos}\label{II}

\subsection{Effective action and de Sitter instanton}

The quantum tunneling can be described in terms of instantons ---solutions to the Euclidean equations of motion.
The effective action $\Gamma$ is defined as
\begin{align}
 e^{-\Gamma}:=\int[{\cal D}g]\,e^{-S_{\rm{eff}}[g]},\,
\end{align}
with the matter effective action $S_{\rm{eff}}$ \emph{defined} by the `quantum average' over matter fields $\Phi(x)$
\begin{align}
 e^{-S_{\rm{eff}}[g]}:=\int\,[{\cal D}\Phi]\,e^{-S[g,\,\Phi]}\,.\label{EffAct1}
\end{align}
In practice, it is usually impossible to calculate the full effective action $\Gamma$ exactly
and one has to resort to a loop expansion. Here, we have split the functional integrals into two parts, 
distinguishing between the geometrical and the matter part. In what follows, we consider the theory described 
by the action $S$ as a quantum field theory in curved spacetime and neglect graviton loops, that is,
we only consider (\ref{EffAct1}) in a classical background described by the metric $g_{\mu\nu}$.

Following \cite{Barvinsky:2009jd}, we consider the matter effective action
\begin{align}
 S_{\rm{eff}}[g]=\int\text{d}^4x\,\sqrt{g}\,\frac{M_{\rm{P}}^2}{2}\left[2\,\Lambda_{\rm{eff}}-R(g)+\dots\right]\,.\label{EHAction1}
\end{align}
Here, $\Lambda_{\rm{eff}}$ is the effective cosmological constant, $M_{\rm{P}}=m_{\rm{P}}/\sqrt{8\,\pi}\approx2.43\times 10^{18}$ GeV is the reduced Planck mass (in units where $\hbar=c=1$) and $R(g)$ is the Ricci scalar constructed from the Euclidean metric field $g_{\mu\nu}(x)$.
The ellipsis stands for higher-order curvature and gradient terms that we do not take into account.

In the cosmological context of slow-roll inflation driven by a real scalar field $\varphi$, the vacuum energy density during inflation is dominated by the nearly constant potential $V(\varphi)$.
This leads to the identification $M_{\rm{P}}^2\,\Lambda_{\rm{eff}}\approx V_{\rm{eff}}(\varphi)$ and justifies the omission of gradient terms in (\ref{EHAction1}).

Once we have calculated the effective action (\ref{EHAction1}), 
we can specialize to a fixed closed Friedmann--Lema\^itre--Robertson--Walker (FLRW) background with line element
\begin{align}
 \text{d}s^2=N^2(\tau)\,\text{d}\tau^2+a^2(\tau)\,\text{d}^2\Omega^{(3)}.\label{LineElFRW}
\end{align}
Here, $a(\tau)$ and $N(\tau)$ are the Euclidean scale factor and lapse function, while $\text{d}^2\Omega^{(3)}$ is the volume element of the three-dimensional sphere.
In the background metric (\ref{LineElFRW}), the effective action (\ref{EHAction1}) reduces to
\begin{align}
S_{\rm{eff}}[a,\,N]=12\,\pi^2\,M_{\rm{P}}^2\,\int\text{d}\tau\,N\,\left[-\frac{1}{N^2}\left(\frac{\text{d}a}{\text{d}\tau}\right)^2a-a+H_{\rm{eff}}^2\,a^3\right]\,,\label{EffActDS}
\end{align}
where we have identified the effective cosmological constant $\Lambda_{\rm{eff}}\equiv3\,H_{\rm{eff}}^2$ 
with the effective Hubble parameter $H_{\rm{eff}}$. 
The instanton is a solution of the Euclidean Friedmann equations, which are obtained by varying (\ref{EffActDS}) with respect to $N$,
\begin{align}
\frac{1}{N^2} \left(\frac{\text{d}a}{\text{d}\tau}\right)^2=1-H_{\rm{eff}}^2\,a^2\,.\label{EEOM1}
\end{align}
This equation has one turning point $a_{+}:=a(\tau_{+}):=1/H_{\rm{eff}}$ so that the real solution interpolates
between $a_{-}:=a(\tau_{-}):=a(0)=0$ and $a_{+}$. Depending on the sign of $N$,
the gauge choice of the Lagrange multiplier $N$ describes two disjoint equivalence classes of instantons.
It is sufficient to consider the representative values $N_{\pm}:=\pm 1$. 
The explicit solution to the differential equation (\ref{EEOM1}) 
then reads
\begin{align}
 a(\tau)=\frac{1}{H_{\rm{eff}}}\text{sin}\left(H_{\rm{eff}}\,\tau\right)\,,\label{ESola1}
\end{align}
where we have fixed the integration constant by the condition
\begin{align}
 \frac{\text{d}\,a}{\text{d}\tau}\Big|_{a=0}=1\,,
\end{align}
which is provided by the constraint equation (\ref{EEOM1}). We have also chosen
the geometrical meaningful positive root of (\ref{EEOM1}) to obtain (\ref{ESola1}).
The turning point $a_{+}$, corresponding to the equator of the Euclidean half sphere where $a(\tau)$ is maximized, determines the `moment of nucleation' $\tau_{+}:=\pi/(2\,H_{\rm{eff}})$.
The tunneling process is then described by attaching the Euclidean half sphere to the inflationary Lorentzian regime at $\tau_{+}$.
At the boundary, we analytically continue (\ref{ESola1}) to Lorentzian signature $\tau\to \rmi\,t$:
\begin{align}
 a_{\rm{L}}(t)=a\left(\frac{\pi}{2\,H_{\rm{eff}}}+\rmi t\right)
 =\frac{1}{H_{\rm{eff}}}\,\text{cosh}\left(H_{\rm{eff}}\,t\right)\,.
\end{align}
The instanton is obtained by inserting (\ref{ESola1}) into (\ref{EffActDS}) and integrating from $\tau_{-}$ to $\tau=\tau_{+}$:
\begin{align}
 S_{\rm{eff}}^{\text{on-shell}}[a,N_{\pm}]=\mp 8\pi^2\,\frac{M_{\rm{P}}^2}{H_{\rm{eff}}^2}\,.\label{EffFRw}
\end{align}
Neglecting graviton loops, we obtain the tunneling instanton $\Gamma^{\text{on-shell}}$ for the choice $N_-=-1$ \cite{Barvinsky:2009jd},
\begin{align}
 \Gamma^{\text{on-shell}}(\varphi):= S_{\rm{eff}}[a,-1]=24\,\pi^2\,\frac{M_{\rm{P}}^4}{V_{\rm{eff}}(\varphi)}\,,\label{EInst}
\end{align}
where we have again used the identification
$H_{\rm{eff}}^2\equiv\Lambda_{\rm{eff}}/3\equiv V_{\rm{eff}}/(3\,M_{\rm{P}}^2)$ in the last step (note that
$V_{\rm{eff}}>0$). 

It should be mentioned that the analogy with the quantum mechanical
tunneling is of at most heuristic value.  
In fact, the presented derivation of the instanton simply corresponds
to a solution of the Euclidean equations of motion subject to some
specially chosen boundary condition. Here, `tunneling' is then
simply \emph{defined} by this choice. In the ordinary quantum
mechanical tunneling problem, 
say the example of the spontaneous decay of an $\alpha$-particle,
tunneling is described by a wave function that contains only outgoing
modes. 
But in this case, there is always a fixed reference phase 
$\propto\text{exp}(-{\rm i}\,\omega\,t)$ with respect to which one can define 
 outgoing and incoming modes unambiguously \cite{Zeh88,Zeh07}. 
The sign in front of the frequency $\omega$ and the external time
parameter $t$ in the exponential is fixed by the sign of the time
derivative in the Schr\"odinger equation. 
If the wave function corresponds to a plane wave
$\propto\text{exp}(-{\rm i}\,\omega t\mp\,{\rm i}\,k\,x)$, a relative
minus sign with respect to the sign of $t$    
corresponds to outgoing modes $k$. In contrast, in the context of the
quantum tunneling of the universe as a whole, there is no such simple
notion of ingoing and outgoing modes, as there is no notion of an
external time parameter anymore at the fundamental level \cite{Kiefer:2012boa}. 
In the absence of any reference phase, the definition of incoming and
outgoing becomes meaningless. The only notion of time one can
introduce at the fundamental level is that of `internal time'. In this
case, the role of time can be played by one or more configuration-space degrees of
freedom. In the context of cosmological minisuperspace, the scale
factor has a preferred role as internal time parameter, in the sense
that its associated kinetic term comes with a relative minus sign
compared to the matter degrees of freedom. This is a consequence of
the indefinite nature of the minisuperspace DeWitt metric. Time as an
external parameter can only be recovered at a semi-classical level 
\cite{Kiefer:2012boa,Kiefer:2013jqa}. All
this does not invalidate the construction presented here but simply
serves to clarify our notion of `tunneling' and emphasizes the difference with respect to the ordinary quantum mechanical tunneling
problem. 

\subsection{Tunneling distribution function and initial conditions for inflation}\label{sec2.2}

The interpretation of the wave function of the universe is largely an open problem \cite{Kiefer:2012boa}.
One heuristic approach is to interpret peaks in the (absolute square of the) wave function as a prediction; see \cite{Hartle:1987}.
Recently, this idea was applied to the model of non-minimal Higgs inflation \cite{Barvinsky:2009jd, Barvinsky:2012zz} and we will follow a similar idea in the present paper with the purpose of presenting a general construction that can serve as a tool to derive predictions from quantum cosmology.

In the semi-classical approximation to quantum cosmology, the no-boundary proposal does usually not give a wave function which is peaked at a field value high enough for inflation \cite{Barvinsky:1990ga}. Such a peak may arise in models of eternal inflation using the landscape picture \cite{Hartle:2010dq} but we will not discuss them here.
For this reason, we will only address the tunneling proposal. From it, using (\ref{EInst}), 
the probability distribution in the semi-classical limit is found to be
\begin{align}
 {\cal T}(\varphi):=e^{-\Gamma^{\text{on-shell}}(\varphi)}=\text{exp}\left[-\frac{24\,\pi^2\,M_{\rm{P}}^4}{V_{\rm{eff}}(\varphi)}\right]\,.\label{ProbDistr1}
\end{align}
A peak corresponds to a maximum of (\ref{ProbDistr1}).
Finding this peak is equivalent to finding the maxima of the potential $V_{\rm{\max}}:=V_{\rm{eff}}(\varphi_{\rm{\max}})$.
This leads to the simple conditions
\begin{align}
 \frac{\text{d}\,V_{\rm{eff}}(\varphi)}{\text{d}\varphi}\Big|_{\varphi=\varphi_{\rm{max}}}=0,\qquad \frac{\text{d}^2\,V_{\rm{eff}}(\varphi)}{\text{d}\varphi^2}\Big|_{\varphi=\varphi_{\rm{max}}}<0\,.\label{MaxCond}
\end{align}
The peak $\varphi_{\rm{max}}$ in (\ref{ProbDistr1}) corresponds to the value of $\varphi$ that selects the most probable value of $\Lambda_{\rm{eff}}=V_{\rm{eff}}(\varphi_{\rm{max}})/M_{\rm{P}}^2$ for which the universe starts after tunneling.
In this way, the quantum scale of inflation was obtained in \cite{Barvinsky:1995gi,Barvinsky:1996ce, Barvinsky:1998qh, Barvinsky:1998rn,Barvinsky:1999qn}.

A high value of $V_{\rm{max}}$ is necessary to start an inflationary evolution after tunneling.
Therefore, $\varphi_{\rm{max}}$ can be interpreted as setting the initial conditions for inflation.
In the inflationary slow-roll regime, $\varphi\approx\text{const}$ and the energy density is completely dominated by $V_{\rm{max}}$.
The peak value $\varphi_{\rm{max}}$ allows one to determine the energy scale of inflation by
\begin{align}
 E_{\rm{infl}}^{\rm{QC}}:=V_{\rm{max}}^{1/4}\,\label{EInfQC}.
\end{align}
One should bear in mind that
$E_{\rm{infl}}^{\rm{QC}}\neq\varphi_{\rm{max}}$ in general, as
$\varphi$ is only a coordinate in the field configuration space and as
such does not have any direct physical meaning  (although both
$E_{\rm{infl}}^{\rm{QC}}$ and $\varphi_{\rm{max}}$ have the same
physical dimension of an energy).
Only the (effective) potential itself can serve as a meaningful observable.

Classical inflationary models predict an energy scale
\begin{align}
 E_{\rm{infl}}^{\rm{model}}:={}& V_{*}^{1/4}\,,\label{EInfModel}
\end{align}
where $V_*:=V(\varphi_*)$ and $\varphi_{*}$ denotes the field value evaluated at the moment $k_{*}=H_{*}\,a_{*}$
when the pivot mode $k_*$ (to be chosen according to the observational window of the experiment) first crosses 
the Hubble scale.
Inflationary models allowing for a quantum cosmological origin in the sense discussed here must therefore satisfy the approximate consistency condition
\begin{align}
 E_{\rm{infl}}^{\rm{QC}}\approx E_{\rm{infl}}^{\rm{model}}\,,\label{ConsistencyCond1}
\end{align}
that is, the energy scale of the inflationary model must be of the
same order as the prediction from quantum cosmology. In principle,
this is an exact relation and one could derive a very precise
prediction of quantum cosmology. However, since in most situations
only a truncated loop expansion of $\Gamma$ (and therefore of ${\cal
  T}$) is available, one cannot expect this condition to
be satisfied exactly at the perturbative level. It is well known that
radiative corrections can change the shape of the effective potential
and, in particular, its extrema which determine the peak position. 

If the amplitude of the tensor power spectrum at the Hubble-scale crossing, $A_{\rm{t}\,*}$,
is known, one can introduce a third scale, the energy scale of inflation inferred 
from this amplitude. 
Using, e.g., the relations (152) and (216) from \cite{BaumannTASI}, one 
gets the following expression for the inferred ``observed'' energy scale of inflation:
\begin{align}
E_{\rm{infl}}^{\rm{obs}}=M_{\rm{P}}\,\left(\frac{3}{2}\,\pi^2\,A_{\rm{t}\,*}\right)^{1/4}
\approx M_{\rm{P}}\,\left(\frac{3}{2}\,\pi^2\,A_{\rm{s}\,*}\,r_{*}\right)^{1/4}.
\end{align}
In the first step, we have used the well-known expression for tensor modes $A_{\rm{t}}\propto H^2\propto E_{\rm{infl}}^4$ at horizon crossing.
To first order in the slow-roll approximation, the scale $E_{\rm{infl}}^{\rm{obs}}$ can thus be expressed in terms of the tensor-to-scalar ratio $r:=
A_{\rm{t}}/A_{\rm{s}}$ and the amplitude of the scalar perturbations
$A_{\rm{s}}$, which is fixed by the measured temperature anisotropies
of the cosmic microwave background. For the pivot scale $k_{*}=
0.002\,\, \text{Mpc}^{-1}$, the best fit of the \textsc{Planck}+WP
data by the $\Lambda$CDM model yields the following $1\sigma$
experimental bound on $A_{\rm{s}\,*}$ \cite{Ade:2013uln}: 
\begin{align}
  \text{ln}\left(10^{10}\,A_{\rm{s}\,*}\right)=3.089^{+0.024}_{-0.027}\,.\label{AsCMB} 
 \end{align}
Until recently, observations gave only an upper bound on $r$. If, however,
the recent announcement by the \textsc{Bicep2} experiment
of the discovery of primordial gravitational waves \cite{Ade:2014xna}
is confirmed, this will yield a model-independent determination of
the energy scale of inflation.
Assuming that this is the case, one
has $r=0.20^{+0.07}_{-0.05}$ for the primordial graviton contribution,
or $r=0.16^{+0.06}_{-0.05}$ if currently best available dust models
are taken into account. Taking the central values
$A_{\rm{s}\,*}=2.2\times10^{-9}$ and $r=0.16$, this leads to an energy
scale 
\begin{align}
 E_{\rm{infl}}^{\rm{obs}}\approx 2.06\times 10^{16} \,\text{GeV}\,,
\end{align}
 which can be roughly taken as upper bound for the inflationary energy if the \textsc{Bicep2} constraint is ignored.
We can thus make contact with experiments via the extended consistency condition
\begin{align}
 E_{\rm{infl}}^{\rm{QC}}\approx E_{\rm{infl}}^{\rm{model}}\approx E_{\rm{infl}}^{\rm{obs}}\,.\label{ConsistencyCond2}
\end{align}

Since probabilistic arguments in the context of cosmology involve
difficult conceptual questions, we shortly summarize the underlying
assumptions allowing for a consistent application and interpretation
of the tunneling scenario presented here. 

\begin{enumerate}
\item \emph{A classical background must have emerged}. In the underlying
  full quantum theory, there is not yet any notion of `background'
  or `classical', but only that of a pure quantum state,
  corresponding to the wave function of the universe.
To understand the semi-classical limit, two steps must be performed 
\cite{Kiefer:2012boa}. First, one must employ a Born--Oppenheimer type
of approximation scheme to find wave functions with a semi-classical
behavior. Second, one must invoke the process of decoherence
\cite{deco} to understand the degree of classical behavior.
  The semi-classical wave function can be understood as
  one branch of the full wave function in the Everett
  interpretation of quantum mechanics. It is this semi-classical
  branch of the full wave function which is used to construct the
  probability distribution (\ref{ProbDistr1}). The emergence of a
  quasi-classical background via decoherence is induced by the
  division of the configuration space into
  system and environment; in concrete models, the system consists of
  global degrees of freedom such as the scale factor and the inflaton,
  and the environment consists of small density fluctuations and small
  gravitational waves \cite{Zeh86,CK87,BKKM97}. 
 The inevitable interaction with the
  environment then leads to the entanglement of the system with the
  environment. The process of decoherence describes the effective
  influence of the environment on the system by integrating out the  
overwhelmingly many inaccessible environmental degrees of freedom and
leads to a suppression of quantum correlations in the reduced density
matrix for the system. The exponential suppression of the non-diagonal
elements of the reduced density matrix then corresponds to an
effective classicalization of the system. Once the classical behavior
of the `background' is understood, one can address the
quantum-to-classical transition of inhomogeneous degrees of freedom
\cite{Kiefer:2006je,Kiefer:2008ku}.

\item \emph{The universe `nucleates' into a homogeneous and isotropic
    universe}. The concordance model of cosmology is based on the
  cosmological principle which implies homogeneity and isotropy around
  any point in space, when averaged over scales larger than around
  $100$ Mpc. This assumption is supported {\it a posteriori} by empirical
  evidence from observations of the large-scale structure within the
  observable patch of our universe.  

\item \emph{Right after the tunneling process, the universe starts a
    phase of accelerated expansion} (inflation). 
This assumption is also supported by observational evidence and {\it a
posteriori} justifies the use of the de Sitter instanton. 

\item \emph{The probability distribution should possess a sharp peak}.
Without such a peak, there would be no clear selection mechanism for
the most probable value of $\varphi$.  If no peak is present and no other criterion is found, one must
refer to the anthropic principle as the only selection principle. 

\item \emph{We have to assume some kind of `principle of
    mediocrity'} \cite{Vilenkin:2011dd} in order to attribute
  predictive power to the result for $E_{\rm{inf}}^{\rm QC}$. This simply means that
  in order to interpret a deviation from the peak of ${\cal T}$ in a
  probabilistic sense, we have to assume that in the multiverse
  context our universe is not very special. Otherwise, a deviation of
  the measured $E_{\rm{inf}}^{\rm{obs}}$ from the calculated
  $E_{\rm{inf}}^{\rm QC}$, determined by the peak of the tunneling
  distribution, would have no predictive power at all, for it might
  perfectly be that we simply live in a very improbable branch of the
  universe located at the far end of the tail of the probability
  distribution, without any contradiction and without being able to
  draw any conclusion from it. In contrast, if we assume instead that
  our semi-classical branch of the universe is indeed for some reason
  mediocre with respect to all other branches, a strong discrepancy  
between the measured $E_{\rm{inf}}^{\rm{obs}}$ and the calculated
$E_{\rm{inf}}^{\rm QC}$ could indicate a falsification of the underlying
inflationary model used to calculate $E_{\rm{inf}}^{\rm QC}$. This
assumption is rather speculative and, of course, tightly related to the inevitable problem of
having only one sample universe. 
\end{enumerate}

\section{Natural inflation}\label{III}

In what follows, we will focus on a tree-level analysis for the model of natural inflation \cite{Freese:1990rb}.
There are several other models favored by \textsc{Planck} data.
Among them, we mention Starobinsky's $R+R^2$ model \cite{Starobinsky:1980te}, inflation with a strong non-minimal coupling \cite{Fakir:1990eg,Spokoiny:1984bd,Salopek:1988qh,Ketov:2012jt,Kallosh:2013tua} (including non-minimal Higgs-inflation \cite{Bezrukov:2007ep,Barvinsky:2008ia,Bezrukov:2008ej,Barvinsky:2009fy,DeSimone:2008ei,Barvinsky:2009ii,Bezrukov:2010jz,Bezrukov:2009db}) and effective string-inspired models (see, e.g., \cite{Dvali:1998pa,Cicoli:2008gp} and related work).
These models predict a tiny tensor-to-scalar ratio which would be in agreement with the upper bound on $r$ derived by \textsc{Planck}, but
in the light of the \textsc{Bicep2} data they are under some pressure. In contrast, the natural-inflation scenario fits the \textsc{Bicep2} data easily \cite{CKT}.

Moreover, while natural inflation already admits a quantum cosmological analysis at the tree level, 
the same is not true for the remaining models listed above. Although the procedure of our tunneling analysis 
is applicable in general also for these models, they all share the common feature that their tree-level potentials 
become nearly flat for high energies and thus do not feature a strict maximum. Hence, there is no sharp peak 
in (\ref{ProbDistr1}). Note, however, that radiative corrections will in general change the structure of the effective potential 
such that a tunneling analysis may become possible.
This was, for example, the case in \cite{Barvinsky:2009jd} where the renormalization-group 
flow of the Higgs potential due to loop contributions of heavy Standard Model particles leads to the formation 
of an additional minimum for high energies, thereby creating a maximum in between the two minima.
We will comment on different models and radiative corrections in section \ref{IV}.  

The potential for natural inflation reads \cite{Freese:1990rb}
\begin{equation}
 V=\Lambda^4\,\left[1+\text{cos}\left(\varphi/f\right)\right]\,.\label{NatInfPot}
\end{equation}
Here, $\varphi$ is interpreted as a pseudo Nambu--Goldstone boson taking values
on a circle with radius $f$ and angle $\varphi/f\in[0,\,2\,\pi)$. the constants $\Lambda$ and $f$
have dimension of mass and determine the height and the slope of the potential; in the model of natural inflation, one expects $f=O(M_{\rm{P}})$ and $\Lambda\approx M_{\rm{GUT}}\sim 10^{16}$ GeV, the grand-unification scale.

\subsection{Tunneling analysis}

The extrema of (\ref{NatInfPot}) are obtained by the condition
\begin{align}
\frac{\text{d} V}{\text{d}\varphi}\Big|_{\varphi=\varphi_{\rm{ext}}}=-\frac{\Lambda^4}{f}\,\text{sin}\left(\varphi_{\rm{ext}}/f\right)=0\,,
\end{align}
leading to $\varphi_{\rm{ext}}=n\,\pi\,f,\; n\in\mathbb{Z}$. If $\varphi_{\rm{ext}}$ is a maximum,
\begin{align}
\frac{\text{d}^2 V}{\text{d}\varphi^2}\Big|_{\varphi=\varphi_{\rm{ext}}}=
-\frac{\Lambda^4}{f^2}\,\text{cos}\left(n\pi\right)<0\,;\label{SecDervPotNaturalTree}
\end{align}
peak values correspond to even $n$, i.e., due to periodicity,
\begin{align}
\varphi_{\rm{max}}:=2\,\pi nf\,.
\end{align}
The potential at $\varphi_{\rm{max}}$ has the value
\begin{align}
V_{\rm{max}}=2\,\Lambda^4\,.\label{VatMax}
\end{align}
The predictability of the tunneling distribution ${\cal T}(\varphi)$ defined in (\ref{ProbDistr1}) 
is determined by the sharpness of the peak $\varphi_{\rm{max}}$. The sharpness is here defined as
\begin{align}
{\cal S}:=\frac{(\Delta\varphi)^2}{\left(E_{\rm{infl}}^{\rm{QC}}\right)^2}
\equiv \frac{(\Delta\varphi)^2}{\sqrt{V_{\rm max}}}\,.
\end{align}
Here, the variance $(\Delta\varphi)^2$ measures the width of the peak, and $E_{\rm{infl}}^{\rm{QC}}$ defines the height of the peak.
In view of (\ref{NatInfPot}), the distribution (\ref{ProbDistr1}) is clearly symmetric around $\varphi_{\rm{max}}$.
In order to get a rough estimate for the width $\Delta\varphi$,
we can fit ${\cal T}$ to a normal distribution around the peak $\varphi_{\max}$. 
Therefore, we take $\varphi_{\rm{max}}$ as the mean and expand
$\Gamma^{\text{on-shell}}$ from (\ref{EInst}) around $\varphi_{\rm{\max}}$ to second order:
\begin{eqnarray}
\Gamma^{\text{on-shell}}(\varphi)&=&\Gamma^{\text{on-shell}}(\varphi_{\rm{\max}})+\frac{1}{2}
\Gamma^{\text{on-shell}''}(\varphi_{\rm{\max}})(\varphi-\varphi_{\rm{\max}})^2\nonumber\\ 
&\equiv&
\Gamma^{\text{on-shell}}(\varphi_{\rm{\max}})+\frac{1}{2}\frac{(\varphi-\varphi_{\rm{\max}})^2}{(\Delta\varphi)^2}.
\end{eqnarray}
This leads to the identification
\begin{align}
(\Delta\varphi)^2:= \left.\frac{1}{\Gamma^{\text{on-shell}''}}\right|_{\varphi=\varphi_{\rm{max}}}=\frac{1}{6\,\pi^2}\,\frac{f^2\,\Lambda^4}{M_{\rm{P}}^4}\,,\label{varianceGaussFit}
\end{align}
where primes denote derivatives with respect to $\varphi$. The sharpness of the peak ${\cal S}$ is then estimated as
\begin{align}
 {\cal S}=\frac{(\Delta\varphi)^2}{\left(E_{\rm{infl}}^{\rm{QC}}\right)^2}
 \approx\frac{1}{6\,\pi^2}\,\frac{f^2\,\Lambda^2}{M_{\rm{P}}^4}\sim\frac{\Lambda^2}{M_{\rm{P}}^2}\sim 10^{-4}\,,
\end{align}
where we have used $f\sim M_{\rm{P}}$ and, using (\ref{EInfQC}) and (\ref{VatMax}),
$E^{\rm{QC}}_{\rm{inf}}\sim\Lambda$.

\subsection{Slow-roll analysis}

The cosmological parameters in the inflationary slow-roll analysis are completely determined by the potential and its derivatives.
The first two slow-roll parameters are given by
\begin{align}\label{srpa}
 \epsilon_{\rm{v}}:={}&\frac{M_{\rm{P}}^2}{2}\,\left(\frac{V'}{V}\right)^2=\frac{M_{\rm{P}}^2}{2\, f^2}\,\text{tan}^2\left[\varphi/(2\, f)\right],\qquad \eta_{\rm{v}}:={}M_{\rm{P}}^2\,\left(\frac{V''}{V}\right)=-\frac{M_{\rm{P}}^2\, \text{cos}(\varphi/f)}{f^2\, \left[1 + \text{cos}(\varphi/f)\right]}\,,
\end{align}
with $\epsilon_{\rm{v}}\ll1$ and $|\eta_{\rm{v}}|\ll 1$ during inflation.
The scalar spectral index and the tensor-to-scalar ratio read
\begin{align}
n_{\rm{s}}={}&1+2\,\eta_{\rm{v}}-6\,\epsilon_{\rm{v}}=-\frac{M_{\rm{P}}^2}{f^2}\,\frac{3-\text{cos}(\varphi/f)}{1+\text{cos}(\varphi/f)}\,,\label{SpectralIndex}\\
r={}&16\,\epsilon_{\rm{v}}=\frac{8\,M_{\rm{P}}^2}{ f^2}\,\text{tan}^2\left[\varphi/(2\, f)\right]\,.\label{TensorToScalarRatio}
\end{align}
All cosmological observables have to be evaluated at $\varphi_{*}$, the field value that corresponds to the moment where the pivot mode $k_{*}$ first crosses the Hubble scale.
The number of e-folds $N_*$, which is a measure of how long inflation lasted, connects the end of inflation $\varphi_{\rm{end}}$ with the value $\varphi_{*}$:
\begin{align}
 N_{*}=\int_{\varphi_{*}}^{\varphi_{\rm end}}\frac{\text{d}\varphi}{M_{P}^2}\,\frac{V}{V'}=\frac{2\,f^2}{M_{\rm{P}}^2}\,\ln\left[\frac{\text{sin}\left(\frac{\varphi_{\rm{end}}}{2\,f}\right)}{\text{sin}\left(\frac{\varphi_{*}}{2\,f}\right)}\right]\,.\label{EFoldsNatural}
\end{align}
The value $\varphi_{\rm{end}}$ that determines the upper integration bound in (\ref{EFoldsNatural}) is defined by the breakdown of the slow-roll approximation at $\epsilon_{\rm{v}}(\varphi_{\rm{end}}):=1$,
\begin{align}
 \varphi_{\rm{end}}=2\, f\, \text{arctan} (\sqrt{2}\, f/M_{\rm{P}})\,.\label{PhiEnd}
\end{align}
Inserting (\ref{PhiEnd}) in (\ref{EFoldsNatural}), solving for $\varphi_{*}$ and parametrizing $f$ 
in units of $M_{\rm{P}}$, we find
\begin{align}
 \varphi_{*}={}&2 M_{\rm{P}}\,\alpha\, \text{arcsin}\left(\frac{\alpha\,e^{-N_{*}/2 \alpha^2}}{\sqrt{
  1/2 + \alpha^2}}\right),\label{phiStar}
\end{align}
where $\alpha:=f/M_{\rm{P}}$.
Evaluating the potential (\ref{NatInfPot}) at $\varphi_{*}$ yields
\begin{align}
 V(\varphi_{*})=2\,\Lambda^4\,\left[1-\delta_{V}(\alpha,N_{*})\right]\,,
\end{align}
where we have defined
\begin{align}
 \delta_{V}(N_{*},\alpha):=\frac{2\,e^{-N_{*}/\alpha^2}\,\alpha^2}{1+2\,\alpha^2}\,.
\end{align}
The consistency condition (\ref{ConsistencyCond1}) implies $V_{\rm{max}}=V(\varphi_{*})$, or $ \delta_{V}=0$. Since $N_{*}$ should be in the range $50\lesssim N_{*}\lesssim 60$, this requires $\alpha=0$ for (\ref{ConsistencyCond1}) to be satisfied exactly.

\begin{figure}[h!]
\begin{center}
\includegraphics[scale=0.8]{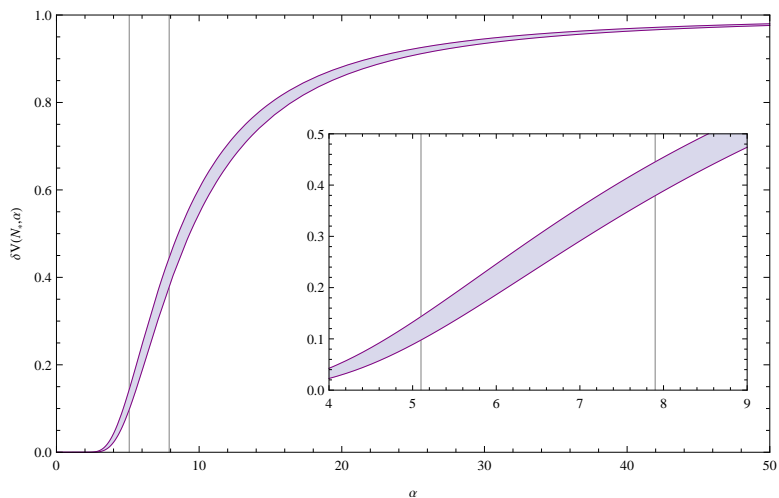}
\caption{\label{fig1} The function $\delta_V(N_{*},\alpha)$ as a function of $\alpha$ for values of $N_{*}\in[50,60]$. The upper line corresponds to $N_{*}=50$, the lower line to $N_{*}=60$. The inset shows the region with $\alpha$ in the $68\%$ CL range $5.1< \alpha < 7.9$ (see (\ref{PlanckDataRange})).
}
\end{center}
\end{figure}

However, by inspection of (\ref{NatInfPot}), $\alpha=0$ is not allowed.
Moreover, for $N_{*}=60$, \textsc{Planck} constraints on $(n_{\rm{s}},\,r)$ \cite{Ade:2013zuv,Ade:2013uln} imply the following bound on $\alpha$ \cite{Tsujikawa:2013ila}:
\begin{align}
\label{alphaPlanck}
 \alpha>4.6\quad(95\%\;\,\text{CL})\,.
\end{align}
For fixed $N_{*}$, 
the function $\delta_{V}(N_{*},\,\alpha)$ varies between zero and one.
As can be seen in figure \ref{fig1}, $\delta_{V}(N_{*},\,\alpha)$ first grows rapidly from zero at $\alpha=0$
to $0.9$ at around $\alpha=20$ and then slowly asymptotes to 1 for $\alpha\to\infty$.

\begin{figure}[ht!]
\begin{center}
\includegraphics[scale=0.9]{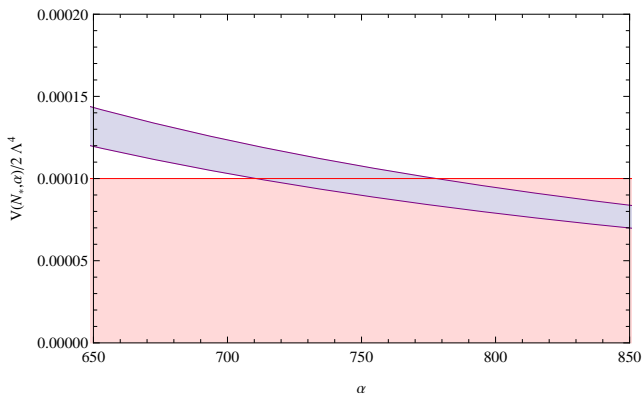}
\caption{\label{fig2} A zoomed-in region of the function $V(N_{*},\alpha)/(2\,\Lambda^4)=1-\delta_V$ as a function of $\alpha$ for values of $N_{*}\in[50,60]$. The upper purple line corresponds to $N_{*}=60$, the lower purple line to $N_{*}=50$. The lower area, colored in light red (in black-and-white printing: light gray), corresponds to the region where $E_{\rm{inf}}^{\rm{model}}<10^{-1}\,E_{\rm{inf}}^{\rm QC}$.
}
\end{center}
\end{figure}

\noindent As already mentioned in the introduction, the consistency condition will lead in general to an exact constraint.
But since we only have considered the tree-level approximation to obtain $E^{\rm{QC}}_{\rm{inf}}$ and also made the
slow-roll approximation to obtain $E_{\rm{inf}}^{\rm{model}}$, this relation cannot be expected to hold exactly. 
Nevertheless, the quantum cosmological analysis leads to an approximate consistency requirement that excludes certain values of $\alpha$ for a given $N_*$.
\noindent Since $\delta_{V}$ enters as the difference $1-\delta_{V}$ in $V(\varphi_{*})$,
it will only lead to significant changes in $E_{\rm{inf}}^{\rm{model}}$ when $\delta_V\approx1$.
For example, a $\delta_{V}\approx 0.9999$ will lead to 
$V(\varphi_{*})=2\,\Lambda^4\left(1-\delta_{V}\right)=2\,\Lambda^4\,10^{-4}$ 
and will affect the energy scale of inflation by one order of magnitude, 
$E_{\rm{inf}}^{\rm{model}}=V^{1/4}(\varphi_{*})=2^{1/4}\times 10^{-1}\,\Lambda$.
Then, even the approximate consistency condition 
$E^{\rm{QC}}_{\rm{inf}}\approx E_{\rm{inf}}^{\rm{model}}$ would no longer hold.
This case would correspond to a value of $\alpha\approx 710$ for $N_{*}=50$ and $\alpha\approx 780$ for $N_{*}=60$ 
and is depicted in figure \ref{fig2}. In order for the consistency condition to hold at least approximately,
we have to impose in this model the constraint
\begin{equation}
 \alpha\ll 700.
 \end{equation}
 This is, of course, compatible with the \textsc{Planck} constraint (\ref{alphaPlanck}).

\begin{figure}[ht!]
\begin{center}
\includegraphics[scale=0.9]{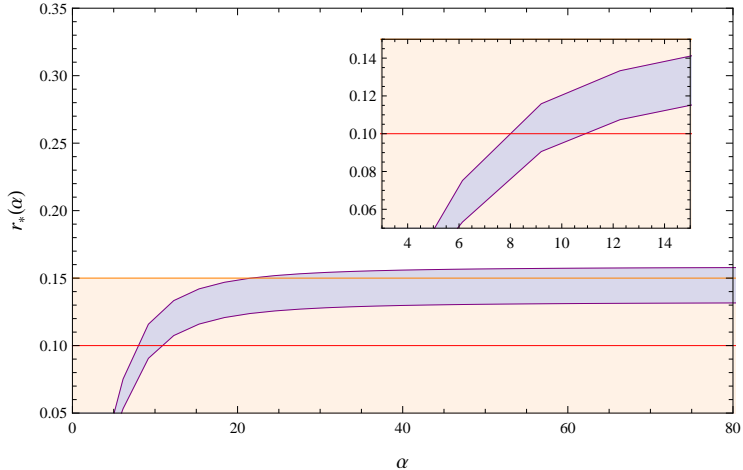}
\caption{\label{fig3} The tensor-to-scalar ratio $r_{*}$ as a function of $\alpha$ for values of $N_{*}\in[50,60]$. 
The $68$ \% CL and $95$ \% CL shaded regions
experimentally excluded by \textsc{Bicep2} are bounded from above by,
respectively, the orange (upper) and red (lower) horizonal line. The upper bounds on $r_{*}$ are not shown,
as they do not constrain $\alpha$. The upper curve corresponds to $N_{*}=50$, the lower one to $N_{*}=60$. 
For $N_{*}\gtrsim 52$, $r_{*}$ falls in the forbidden $68$ \% CL region for all values of $\alpha$.
The inset shows the intersection of the $N_{*}\in[50,60]$ band with the $95$ \% CL bound.}
\end{center}
\end{figure}

Although a quantum cosmological bound on $\alpha$ derived in this way is rather arbitrary and clearly depends
on the aimed precision for the approximate quantum cosmological consistency condition to hold, it is obvious
that this bound is not as restrictive as the constraints on $\alpha$ coming from the comparison of the
inflationary model itself with observational data, which require $\alpha\sim O(10)$.
As described in section \ref{sec2.2}, the observed value $E_{\rm{inf}}^{\rm{obs}}$ is very close to the quantum
cosmological predicted value of $E_{\rm{inf}}^{\rm{QC}}\approx2^{1/4}\,\Lambda$ with $\Lambda\approx10^{16}$ GeV 
around the GUT scale. However, the observational constraints on the spectral index and the tensor-to-scalar ratio by \textsc{Planck} yield a much sharper condition on $\alpha$ than the restrictions from quantum cosmology.
For $N_{*}=60$, \textsc{Planck} data \cite{Ade:2013zuv,Ade:2013uln} constrain $\alpha$ to lie in the interval \cite{Tsujikawa:2013ila}
\begin{align}
 5.1<\alpha<7.9\quad(68\%\,\text{CL})\,.\label{PlanckDataRange}
\end{align}
As shown in figure \ref{fig1}, in this range $\delta_V\approx 0.1\div 0.5$ is small enough to respect the condition (\ref{ConsistencyCond2}), at least within the order of magnitude of the envisaged accuracy.

For the `classical' natural inflation model, a quick estimate for the constraint analogous
to (\ref{PlanckDataRange}) can be obtained by looking at the intersection points in 
the $(n_{\rm{s}\,*},\alpha)$ and $(r_*,\alpha)$ planes between the experimental $68$ \% CL and $95$ \% CL
bounds on $n_{\rm{s}\,*}$ and $r_{*}$ and the corresponding model-dependent analytic expressions 
for $n_{\rm{s}\,*}$ and $r_{*}$. The expressions for $n_{\rm{s}}$ and $r$ given in 
(\ref{SpectralIndex})--(\ref{TensorToScalarRatio}) and evaluated at $\varphi_{*}$ take a particularly simple form when expressed in terms of $\delta_V$ and $\alpha$:
\begin{align}
n_{\rm{s}\,*}={}1+\frac{1}{\alpha^2}\,\frac{\delta_V(N_{*},\,\alpha)+1}{\delta_V(N_{*},\alpha)-1}\,,\qquad r_{*}={}\frac{8}{\alpha^2}\,\frac{\,\delta_V(N_{*},\,\alpha)}{1-\,\delta_V(N_{*},\,\alpha)}\,.
\end{align}
Here we show the results of this procedure by confronting the analytically obtained $r_{*}$ with the estimated bounds on $r_{*}$ from the combined \textsc{Planck}+\textsc{WP}+highL+\textsc{Bicep2} 1$\sigma$ and 2$\sigma$ contours of the tensor-to-scalar ratio for fixed central value $n_{s\,*}=0.96$ \cite{Ade:2014xna}. 
The 1$\sigma$ contour leads to the bounds $r_{*}=0.20^{+0.07}_{-0.05}$ for the tensor-to-scalar ratio. Comparison with figure \ref{fig3} leads to the lower bound $\alpha\gtrsim 22$ for $N_{*}=50$.
The 2$\sigma$ contour of $r_{*}$ for the central value $n_{s\,*}=0.96$ roughly yields the bounds $r_{*}=0.20^{+0.1}_{-0.1}$ \cite{Ade:2014xna}, constraining $\alpha\gtrsim 8$ for $N_{*}=50$ and $\alpha\gtrsim 11$ for $N_{*}=60$, as can be seen in the inset of figure \ref{fig3}. A more elaborate likelihood analysis refines these rough estimates \cite{CKT}:
at the 1$\sigma$ level, $\alpha\gtrsim 9$ for $N_{*}=50$, while at the 2$\sigma$ level $\alpha\gtrsim 7\div 8$ for $N_*=50$ and $\alpha\gtrsim 9$ for $N_*=60$.

Thus, by comparing the classical inflationary predictions with observational data we have obtained a constraint on $\alpha$ of the order of magnitude $\alpha\sim O(10)$ far below the threshold $\alpha\approx 700$ at which a conflict with the quantum cosmological compatibility constraint would arise. We can therefore conclude that, to a good approximation, the consistency condition is satisfied for all experimentally allowed values of $\alpha$ according to both \textsc{Planck} and \textsc{Bicep2} data.

\section{Conclusions}\label{IV}

The purpose of our paper is to present a general method to discussing quantum cosmological consistency conditions 
for inflation and to present a concrete example in detail. We have focused on the tunneling condition for the wave function
of the universe because it allows one to implement these consistency conditions in a straightforward
manner. In principle, however, the method can also be used to study other conditions such as the no-boundary condition,
although this condition does not lead to the prediction of inflation in the usual situations. 
A central concept in our analysis is the use of the effective action. Because this action can
in general not be evaluated exactly,
it is necessary to perform a loop expansion. The example we have discussed in detail here is natural inflation.
There, the restriction to the tree-level approximation seems sufficient because at tree level
the potential (\ref{NatInfPot}) features a strict maximum necessary 
for the tunneling analysis, while this is not the case for the inflationary models mentioned at the beginning of 
section \ref{III}. For these models, therefore, one has to go at least to the one-loop level. This will be
the topic of future investigations.

All inflationary single-field models favored by recent \textsc{Planck} data can be collectively covered by the class of scalar-tensor theories with the action
\begin{align}
 S=\int\text{d}^4x\,\sqrt{|g|}\,\left[U(\varphi)\,R-\frac{G(\varphi)}{2}\,(\nabla\varphi)^2-V(\varphi)\right]\label{ActionScalarTensor1}\,,
\end{align}
where $U(\varphi)$, $G(\varphi)$ and $V(\varphi)$ are arbitrary functions of the inflaton field $\varphi$.
Quantum corrections usually modify the shape of the effective potential and the location of its extrema. Thus, even for the inflationary models where a tunneling analysis was not applicable at the tree level, already at the one-loop level the changing structure of the effective potential may lead to a strict maximum such that a tunneling analysis becomes possible.

The one-loop divergences for the action (\ref{ActionScalarTensor1}), necessary for the renormalization of (\ref{ActionScalarTensor1}), can be extracted from \cite{Steinwachs:2011zs} where a more general action with a $O(N)$-symmetric scalar multiplet was considered.\footnote{A similar analysis has been performed in \cite{Shapiro:1995yc} for a single scalar field. For the divergences that can be absorbed in the functional couplings $U$, $G$ and $V$, the results of \cite{Shapiro:1995yc} coincide with those derived in \cite{Steinwachs:2011zs}.}
However, two important points have to be taken into account for such a quantum analysis.
First, if the inflaton field is coupled to additional matter, matter loop contributions usually lead to a significant modification of the effective potential. This fact was crucial in the renormalization-group improved investigation of the tunneling scenario for non-minimal Higgs inflation \cite{Barvinsky:2009jd}.
Second, the analysis of $f(R)$ theories and models with a non-minimal coupling to gravity requires extra care.
The tunneling formalism presented here has been developed for a minimally coupled scalar field.
By a conformal transformation of the metric field and a subsequent redefinition of the scalar field, the action (\ref{ActionScalarTensor1}) can be brought to the so-called Einstein-frame parametrization, which, in the absence of matter, formally resembles the situation of a scalar field minimally coupled to gravity.
It is well known that $f(R)$ theories with $f_{,RR}\neq0$ also admit an on-shell reformulation as scalar-tensor theories of the type (\ref{ActionScalarTensor1}) with $G(\varphi)=0$.
Therefore, they can ultimately be cast as well in the Einstein-frame parametrization.
While field reparametrizations lead to equivalent descriptions at the tree level, quantum divergences induce a frame dependence of the off-shell effective action \cite{Kamenshchik:2014waa}. In \cite{Vilkovisky:1984st}, the origin of this parametrization dependence was traced back to the non-covariant definition of the off-shell effective action on configuration space and in \cite{Steinwachs:2013tr,Kamenshchik:2014waa} this idea was applied to the cosmological context. In particular, when applied to the debate `Jordan frame vs. Einstein frame', it was pointed out that within a non-covariant formalism quantum corrections will naturally induce a frame dependence when the conformal transformation of the metric field as well as the transformation of the scalar field are viewed as field reparametrizations in configuration space. The tunneling consistency condition may thus serve not only as a tool to distinguish between competing models of inflation, but also to select a preferred parametrization \emph{in the absence} of a covariant formulation.

In our tree-level analysis of natural inflation, we have derived a consistency condition which restricted the parameter $\alpha=f/M_{\rm{P}}$ for a given number of e-folds $N_{*}$.~Our result ensures consistency with the quantum cosmological tunneling origin. For the tree-level analysis of the natural inflation model,
the restriction of $\alpha$ from the quantum cosmological consistency condition
is much weaker than the observational constraints on the inflationary parameters itself.
We have found that natural inflation with a quantum tunneling origin
is consistent with the 2013 \textsc{Planck} release as well as with a large tensor-to-scalar ratio as found by \textsc{Bicep2}. Making use of the general formalism presented here for the natural inflation model, the analysis can easily be extended to all kind of inflationary models, including their modification by quantum corrections. Such investigations will shed further light on the relation between a fundamental theory of quantum cosmology and cosmological observations.

\acknowledgments{The work of G. C. is under a Ram\'on y Cajal contract. C. K. thanks the Max Planck Institute for Gravitational Physics (Albert Einstein Institute) in Potsdam, Germany, for kind hospitality while part of this work was done. C. S. is grateful to A. Yu. Kamenshchik for fruitful discussions and valuable comments. }

\end{document}